\documentclass[conference,10pt]{IEEEtran}
\usepackage[utf8]{inputenc}
\usepackage{color,xcolor}
\usepackage{epsfig}
\usepackage{amsmath,amsfonts,amsthm,amssymb}
\usepackage{graphics,graphicx,pgfplots,tikz,subfigure}
\usepackage{float}
\usepackage{algorithm,algorithmic}
\usepackage{svg}
\usepackage{epstopdf}
\usepackage{soul,csquotes,ulem}
\usepackage[hidelinks]{hyperref}

\begin{document}
\title{Probabilistic Ray-Tracing Aided Positioning at mmWave frequencies}
	\author{
		\IEEEauthorblockN{
			Viet-Hoa Nguyen, Vincent Corlay, Nicolas Gresset, and Cristina Ciochina 
		}
		\IEEEauthorblockA{
			Mitsubishi Electric R\&D Centre Europe, Rennes, France\\
			Email: \{v.nguyen, v.corlay, n.gresset, c.ciochina\}@fr.merce.mee.com
		}
	}
	\maketitle
	
\begin{abstract}
We consider the following positioning problem where several  base stations (BS) try to locate a user equipment (UE): 
The UE sends a positioning signal to several BS. 
Each BS performs Angle of Arrival (AoA) measurements on the received signal.
These AoA measurements as well as a 3D model of the environment are then used to locate the UE.
We propose a method to exploit not only the geometrical characteristics of the environment by a ray-tracing simulation, but also the statistical characteristics of the measurements to enhance the positioning accuracy. 
\end{abstract}



\section{Introduction}

Standard positioning systems rely on line-of-sight (LoS) measurements. 
As an example, if several BS estimate the AoA of a reference signal transmitted by a UE, the regular triangulation method can be implemented as follows. A server, knowing the position of the BS and the measured AoA, looks for the intersection of the rays leaving the BS in the measured AoA directions. The estimated position of the UE is this intersection.

Nevertheless, the received signals at the BS may be non line-of-sight (NLoS). It means that the paths followed by the signals have at least one reflection. In this case, the measured AoA does not indicate the direction of the UE but the direction of the last reflecting object. As a result, the above standard triangulation method is not sufficient for the positioning operation.
One solution to address this issue is to use the knowledge of the environment to improve the positioning operation. 

One can for instance, in an offline phase, create a database of measurements (at the BS) corresponding to each possible position of the UE. Then, in the online phase, a measurement is performed and the most similar measurement in the database is searched. Once found, the position corresponding to the measurement is the estimated position. Similarly, neural network can be trained with the database to infer the UE position based on measurements. This class of methods is called fingerprinting. The database can be established via real measurements or by using a digital twin of the environment combined with ray tracing. This latter approach is for instance considered in \cite{Wielandt2017}\cite{Kikuchi2006}. It is also currently studied at the 3rd Generation Partnership Project (3GPP) Radio Access Network (RAN) 1, in the scope of release 18, within the study item ``AI/ML for positioning accuracy enhancement" \cite{TdocSumOct}.

Alternatively, instead of using a database, online ray tracing can be considered: Given measured AoA, ray tracing in these AoA directions is performed. The intersection of the rays is then the estimated position. This approach, which can be called “reverse ray tracing”, is for instance presented in \cite{Kong2006}. A similar method with the notion of “virtual BS” is described in \cite{Liu2014}. In this latter work, geometric equations rather than conventional ray tracing are considered.

Of course, the measurements may not be perfectly accurate, and the measured AoA may be corrupted by some noise. In \cite{Kong2006}, the authors propose to launch several rays in an interval around the measured AoA. The considered width of the interval is twice the estimated standard deviation of the estimated AoA value. A least square solution, assuming i.i.d. Gaussian noise on the estimated AoA, is used in \cite{Liu2014}.

To the best of our knowledge, the statistics of the measurements are not taken into account in a Bayesian manner in conventional ray-tracing aided positioning. Consequently, we propose to modify the conventional reverse ray-tracing method to take into account the AoA statistics.

\section{Problem statement and statistical modelling}

\subsection{Problem statement}

\begin{figure}[t]
\hspace{-4mm}
    \centering
    \includegraphics[scale=0.59]{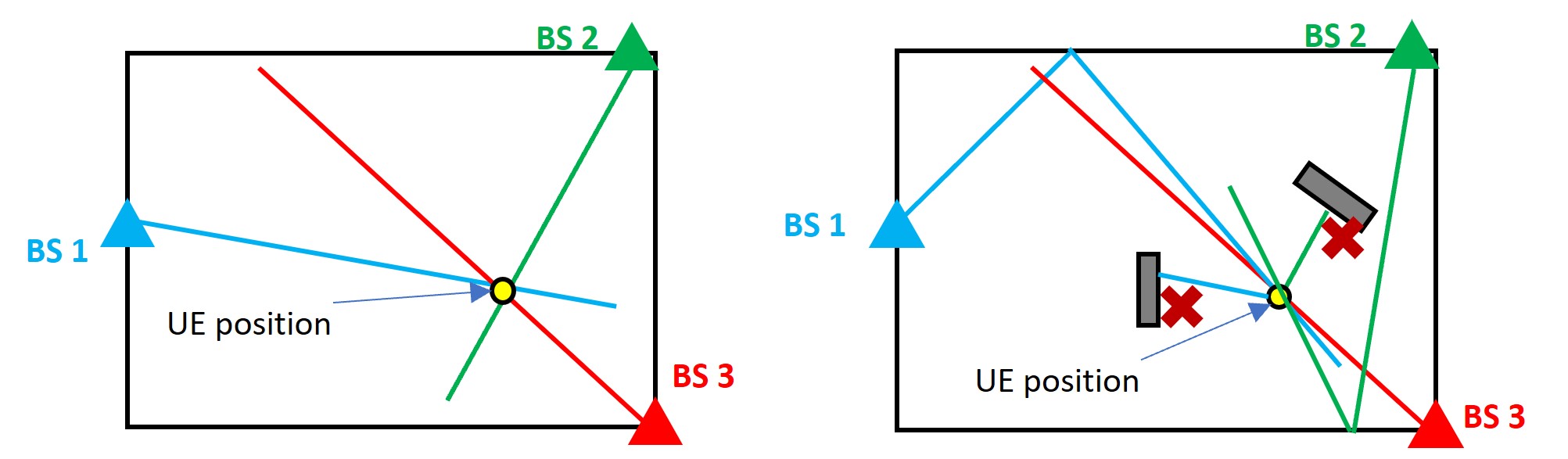}
    \caption{Left: Standard triangulation approach via  LoS-based AoA at the BS. Right: Situation where some LoS AoA are not available for several BS. In the example, two of the three LoS paths are blocked by clutters. In this case,  reverse ray tracing using a digital twin can be considered. }
    \label{fig:example_triang_reverse_ray}
\end{figure}

As mentioned in the introduction, this paper copes with the UE positioning problem, based on the AoA measurements. The UE broadcasts a radio signal and several BS, with fixed and known positions, receive the signal and measure the AoA. In this considered AoA-based positioning framework, two following situations occur:
\begin{itemize}
\item LoS situation:
If all UE-BS links are LoS, the UE lies on the line that crosses the BS with an angle equal to the AoA. Therefore, the intersection of all said lines is the UE position. The standard triangulation technique can be used as illustrated on Figure~\ref{fig:example_triang_reverse_ray} (left).
\item NLoS situation:
If a UE-BS link is NLoS, the UE does not lie on the line that crosses this BS with an angle equal to the AoA. An example is provided on Figure~\ref{fig:example_triang_reverse_ray} (right). In this case, the standard AoA positioning algorithm cannot solve the positioning problem. 
\end{itemize}

In this paper, we focus on indoor positioning scenarios where mostly NLoS situations are encountered. Indeed, in a typical indoor environments, there are many objects of different sizes, shapes, materials, etc. Therefore, the UE is rarely in a LoS situation with the BS. Nevertheless, it is sometimes possible to have a 3D map of the considered indoor environment. This motivates the proposed approach. 

\subsection{Statistical modelling}

Let $X$ be a random variable representing the position of a UE. Let $\Theta$ be a random variable representing the true AoA of the signal, and $Y$ a random variable representing the BS measurement(s) of the AoA. 

The connection between $Y$ and $\Theta$ depends on the antenna hardware and software used by the BS. The BS can have different software and hardware configurations for the AoA measurement. In terms of software, different angle estimation algorithms such as Delay-and-Sum \cite{DS}, MUSIC \cite{MUSIC}, MVDR \cite{MVDR}, ESPRIT \cite{ESPRIT},  induce different error profiles. On the other hand, a distinct hardware configuration, such as the antenna type, the number of antennas, the RF filter, also imposes a particular accuracy of AoA measurement. Hence, each BS may have a specific error model.

For instance, with the standard Gaussian model we have:
\begin{align}
\label{equ_gaus_model}
Y=\ \Theta+W,
\end{align}
where $W\sim\mathcal{N}\left(0,\sigma^2\right)$. The value of $\sigma^2$ may for example depend on the antenna quality. Moreover, if we assume that $\Theta$ is equiprobable, then $p\left(\Theta=\theta\middle| Y=y\right) \propto p\left(Y=y\middle|\Theta=\theta\right) \sim\mathcal{N}\left(\theta,\sigma^2\right)$, where $\propto$ means “proportional to”. Note that $\Theta$ may not always be equiprobable, especially if the BS are located in corners.

The possible values $\theta$ for $\Theta$ is denoted by $\mathcal{C}$. For instance, $\mathcal{C}=[0,2\pi[$ in the 2D case or $[0,\ 2\pi [ \times [0,\ \pi[$ in the 3D case. Note that the interval can be discretized if needed.

Regarding the notations, we use $p\left(\theta\middle| y\right)$ for $p\left(\Theta=\theta|Y=y\right)$ and similarly $p\left(x\middle| y\right)$ for $p\left(X=x|Y=y\right)$.

We let $n$ be the number of measurements such that $\textbf{y} = [y_1,y_2,\ldots,y_i,\ldots,y_n]$.  
The vector $\textbf{y}$ comprises the measurement performed by all BS. For the sake of simplicity, we assume that there is one measurement per BS, and therefore $n$ BS.

In the considered problem, the goal is to compute 
\begin{align}
\label{equ_posi_main}
p\left(x|\textbf{y}\right).
\end{align}
The estimated position $\hat{x}$ is then computed from these estimated probabilities. It can be either estimated by finding the arg max:
\begin{align}
\hat{x}=\text{arg\ max}_x\ p\left(x|\textbf{y}\right), 
\end{align}
or the mean 
\begin{align}
\hat{x}=\int_x  x \cdot p\left(x|\textbf{y}\right) dx. 
\end{align}

\section{Proposed Bayesian approach with ray tracing}

\subsection{Standard Bayesian equation}

In this section, we derive the Bayesian equation to take into account the AoA statistics (i.e., the $p(\theta|y_i)$) in the estimation of $p(x|\textbf{y})$. We let $\boldsymbol{\theta}= [\theta_1,\theta_2,\ldots,\theta_i,\ldots,\theta_n]$.
This simply consists in marginalizing with respect to $\boldsymbol{\theta}$.

Coming back to \eqref{equ_posi_main}, with the law of total probability we have 

\begin{align}
\label{eq_posi_margi}
\begin{split}
p\left(x\middle| \textbf{y}\right)& = \int_{\boldsymbol{\theta} \in\mathcal{C}^n}p\left(x\middle|\boldsymbol{\theta},\textbf{y} \right)  p\left(\boldsymbol{\theta}\middle| \textbf{y} \right) d \boldsymbol{\theta}, \\
&=\int_{\boldsymbol{\theta}\in\mathcal{C}^n} p\left(x\middle|\boldsymbol{\theta} \right)   p\left(\boldsymbol{\theta}\middle| \textbf{y} \right)d \boldsymbol{\theta}.
\end{split}
\end{align}

As the AoA measurement errors of the BS are independent, the probability $p(\boldsymbol{\theta} |\mathbf{y})$ is the product of  the $p(\theta_i|y_i)$: 

\begin{align}
\label{eq_baye}
\int_{\boldsymbol{\theta}\in\mathcal{C}^n} p\left(x\middle|\boldsymbol{\theta} \right)   p\left(\boldsymbol{\theta}\middle| \textbf{y} \right) d \boldsymbol{\theta}=\int_{\boldsymbol{\theta}\in\mathcal{C}^n} p\left(x\middle|\boldsymbol{\theta}\right)\prod_{i=1}^{n}{p\left(\theta_i\middle| y_i\right)}d \boldsymbol{\theta}.
\end{align}

\subsection{Ray tracing}

The aim of this section is to explain the above equation in a NLoS situation using ray tracing.
Indeed, in the LoS situation the term $p\left(x\middle|\boldsymbol{\theta}\right)$ is trivial to compute via the standard triangulation method.
 
We first introduce the notion of ray tracing.

\subsubsection{Reverse ray tracing}
Let us formally introduce the notion of ray. In this paper, a ray is defined as a path, characterized by a 3D scene, starting from a given location and with an angle of departure (AoD)\footnote{The AoA of the received signal becomes the AoD of the ray in reverse ray tracing.}~$\theta$. We use the notation path($\theta$) to refer to a ray. 

If only NLoS AoA measurements are available, a digital twin is required to find the paths of the rays and thus their intersections. This approach is called reverse ray tracing and is illustrated on Figure~\ref{fig:example_triang_reverse_ray} (right). The term ``reverse" is used because the rays are launched from the receiver (the BS who received the UE signal). This approach holds as the considered channel is reciprocal, as justified by the following paragraph.

Regarding the propagation of the rays, we make the following geometric assumption. Due to smaller wavelength, the diffraction phenomenon is less important at millimeter-wave (mmWave) frequencies compared to centimeter-wave frequencies \cite{Deng2016}. Consequently, many channel models consider only specular reflections at the former higher frequencies \cite{Lecci2020}. 
As a result, unlike sub-6 GHz signals, the propagation of mmWave in specific environments can be studied by ray-tracing simulations, as done e.g., in \cite{Degli2014}\cite{Larew2013}\cite{Lecci2020}.


\subsubsection{Interpretation of Equation~\eqref{eq_baye}}
First, the term $p\left(\theta_i\middle| y_i\right)$ corresponds to the statistics of the measurement error of the AoA.
It also represents the distribution of the rays as it enables to compute the probability of a given ray path($\theta_i$).

Second, the term $p\left(x\middle|\boldsymbol{\theta} \right)$ is given by the ray-tracing model. 
Assuming a perfect knowledge of the scene and an unlimited number of rays, $p(x|\boldsymbol{\theta})$ becomes an indicator function: 
 \begin{align}\begin{split} \label{eq_proba_1} &p\left(x\middle|\boldsymbol{\theta}\right)= \\
	&1\{\text{If the }n\text{ rays defined by }\boldsymbol{\theta} = [\theta_1,\theta_2, ...,\theta_n] \text{  cross at }x \}. \end{split} \end{align} 
	 
 \begin{align*}\begin{split} p\left(x\middle|\boldsymbol{\theta}\right)= 0\{ \text{If the rays do not cross at x}\}.   \end{split} \end{align*} 

However, the model for $p\left(x\middle|\boldsymbol{\theta}\right)$ may not be perfect and the number of rays may be finite. As a result, the rays may never cross at one exact location.
In this case, a simple solution is to consider squares (or cubes in the 3D case) $\mathcal{X}$ and to modify \eqref{eq_baye} and \eqref{eq_proba_1} into \eqref{eq_proba_prod} and \eqref{eq_big_square}, respectively, as follows. Let $\mathcal{C}_i'$ be the set of AoD chosen to launch the rays from the $i$-th BS. We get:

\begin{align}
\label{eq_proba_prod}
p\left(\mathcal{X}\middle| \boldsymbol{y}\right) \approx \sum_{\boldsymbol{\theta}\in\mathcal{C}_1'\times \mathcal{C}_2' \times ...}{p\left(\mathcal{X}\middle|\boldsymbol{\theta} \right)\prod_{i=1}^{n}{p\left(\theta_i\middle| y_i\right)}},
\end{align}
where:
\begin{align}
\label{eq_big_square}
\begin{split}
&p\left(\mathcal{X}\middle|\boldsymbol{\theta} \right)= \\
&1\{ \text{If all } n \text{ rays defined by } \boldsymbol{\theta} = [\theta_1,\theta_2,\ ...,\theta_n] \text{ go through } \mathcal{X}\}
\end{split}
\end{align}
 \begin{align*} \begin{split} p\left(\mathcal{X}\middle|\boldsymbol{\theta}\right)=0\{ \text{If not all rays go through } \mathcal{X}\}. \end{split} \end{align*}


Note that the digital twin could also be used to estimate the distribution of $\Theta$.
However, for the sake of simplicity, we assume that it is equiprobable in our simulations. 

\section{Proposed algorithms}

A first natural approach to compute \eqref{eq_proba_prod} via ray tracing is as follows. First, we let $\mathcal{C}'_i$, for all $i$, be a uniformly discretized version of the possible AoA $\mathcal{C}$. Then, for each BS a ray is launched in each resulting direction. Finally, the probabilities are computed via \eqref{eq_proba_prod}, where each ray is weighted by the probability $p(\theta_i|y_i)$. A pseudo code of this approach is provided in Appendix~\ref{App_1}.


%


An alternative manner to compute \eqref{eq_proba_prod} is via Monte Carlo sampling. The main idea is to sample the angles   from the distribution $p\left(\theta_i\middle| y_i\right)$ and launch the rays accordingly. Hence, $\mathcal{C}_i'$ is the set of sampled angles at the $i$-th BS. Then, \eqref{eq_proba_prod} becomes:

\begin{align}
p\left(\mathcal{X}\middle| \boldsymbol{y}\right) \approx \sum_{\boldsymbol{\theta}\in\mathcal{C}_1'\times \mathcal{C}_2' \times ...}{p\left(\mathcal{X}\middle|\boldsymbol{\theta} \right)}.
\end{align}
In other words, once the rays are launched, we simply count the number of times a set of rays crosses the square $\mathcal{X}$ of interest.
The pseudo code for this approach is provided in Algorithm~\ref{Second_alg}.

In the case of a limited number of rays, the Monte Carlo approach is more adapted than the uniform approach (Algorithm~\ref{first_alg}): More samples (i.e., rays) are available at the likely AoA with the former approach than with the latter one. Consequently, one is more likely to have rays crossing in the squares corresponding to high position probabilities. 
This approach is chosen for the simulations.

\begin{algorithm}
\caption{Monte Carlo method to estimate the position via reverse ray tracing}
\label{Second_alg}
\textbf{Inputs:}  AoA measurements $\textbf{y}$, AoA statistics $p(\boldsymbol{\theta}|\textbf{y})$, and 3D model of environment (digital twin). \\
\begin{algorithmic}[1]
\STATE 	For each of the $n$ BS, sample $l$ angles \\ according to $p\left(\theta | y_i\right)$.
\STATE 	Launch the $n\ \times l$ rays in the sampled angle directions.
\FOR{ each distinct $\mathcal{X}_k$ in the scene}
\STATE 	Collect all the rays crossing $\mathcal{X}_k$.
\STATE	Find $m$, the number possible combinations of $n$ rays (where each of the $n$ rays comes from a different BS). Set $\beta_k=m$ to obtain \eqref{eq_proba_prod}.
\ENDFOR
\STATE 	The square with the estimated highest probability is $\mathcal{X}_k$ where ${k=\text{arg \ max}_k\beta}_k$.
\end{algorithmic}
\end{algorithm}


\section{Simulation results}

\subsection{Simulation environment}

\begin{figure}[t]
    \centering
    \includegraphics[scale=0.6]{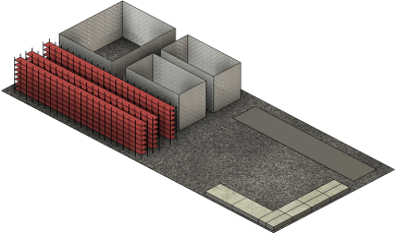}
    \caption{Considered scene for the simulations.}
    \label{fig:scene_digi-twin}
\end{figure}

We use our own Matlab-implemented ray-launching software for the simulation.
The considered scene is shown in Figure~\ref{fig:scene_digi-twin}. The dimensions are the following: width 8m, length 18m, and height 2.5m. 
The scene was drawn via the software Fusion 360. 
This environment is inspired from the recommendations of the 5G Alliance for Connected Industries and Automation (5G-ACIA) for indoor industrial scenario \cite{5G_ACIA}.
It includes both an open area and alleys.

\begin{figure}[t]
    \centering
    \includegraphics[scale=0.24]{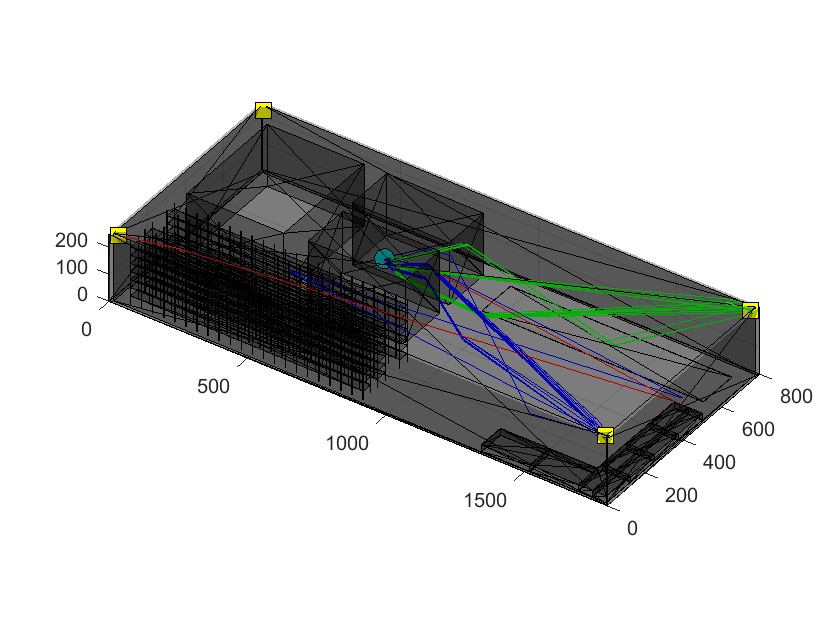}
\vspace{-4mm}
    \caption{Example of ray tracing between the BS and the UE in the considered scene. The positions of the BS are shown by the yellow square and the one of the UE by the blue ball.}
    \label{fig:scene_digi-twin_rays}
\end{figure}

Figure~\ref{fig:scene_digi-twin_rays} shows an example of rays launched from a given position. 
Only the rays crossing both the UE (cyan circle) and at least one BS (yellow squares) are shown. The maximum number of bounces is set to 5.  This value is high enough as the radio signal loses a lot of energy after each bounce at mmWave frequencies.
The 4 BS, located in the top corners of the scene, are shown by the yellow squares. We see that one BS receives no ray.

\subsection{Noiseless AoA and error model}

Since we do not have a real deployment of the scene, the noiseless AoA resulting from the transmission between the UE and the BS must also be established in the digital twin.
The AoA for a given position is established as follows: Many rays are launched from the given position. Only a subset of these rays crosses a given BS. 
We then perform binning of the AoA of the crossing rays. Finally, the angle corresponding to the center of the bin having the highest number of crossing rays is chosen as the AoA for the considered position. This process has to be performed in an offline manner for every considered position.

Regarding the error model, we consider a simple Gaussian noise. Let $\theta_i^k$ be the established AoA for a $i$-th BS and a UE position $x_j$.
The measurement $y^j_i$ is obtained as in \eqref{equ_gaus_model}: $y^j_i=\theta^j_i +w$ where $w$ is sampled from a centered Gaussian distribution with variance $\sigma^2$. 
Hence, the distribution $p(\theta_i | y_i)$ is also Gaussian.



\begin{algorithm}
\caption{Benchmark reverse ray tracing algorithm}
\label{third_alg}
\textbf{Inputs:} AoA measurements $\textbf{y}$, Noise variance $\sigma^2$, and 3D model of environment (digital twin). \\
\begin{algorithmic}[1]
\STATE 	For each of the $n$ BS, generate $l$ angles in the range $[-\sigma,+\sigma]$ discretized uniformly.
\STATE 	Launch the $n\ \times l$ rays in the according to the generated angles.
\FOR{ each distinct $\mathcal{X}_k$ in the scene}
\STATE 	Collect all the rays crossing $\mathcal{X}_k$.
\STATE	Find $m$, the number possible combinations of $n$ rays (where each of the $n$ rays comes from a different BS). Set $\beta_k=m$. 
\ENDFOR
\STATE 	The square with the estimated highest probability is $\mathcal{X}_k$ where ${k=\text{argmax}_k\beta}_k$. 
\end{algorithmic}
\end{algorithm}

\subsection{Benchmark algorithm}

We compare our algorithm with a benchmark algorithm similar to the one proposed in \cite{Kong2006} (mentioned in the introduction). 
With this benchmark algorithm, the rays are launched uniformly in a cone whose width depends on the variance of the error but where the rays are not weighted by probabilities. It is summarized in Algorithm~\ref{third_alg}.

Note that once the rays are launched, the counting is done as in Algorithm~\ref{Second_alg}. 
Hence, the only difference between these two algorithms lies in the way the rays are launched. 
The complexity of both methods is therefore the same.

\subsection{Simulation parameters}

We consider only the 2D $(x,y)$ location problem. 
The height of the UE to locate is assumed fixed (and known) at 1m above the floor. Several random UE positions are considered.

The size of the squares $\mathcal{X}$ should be neither too large nor too small: In the former case, the positioning accuracy would be reduced as we have no indication of the position within a square. In the latter case, if the number of rays is not large enough, we may have no square where at least $n$ rays cross, as required by \eqref{eq_big_square} to have a non-zero probability. As a result, there is a trade-off to find. Of course, the larger the number of rays, the smaller the size of square and the better the performance, but also the higher the complexity.
In our simulations, we choose squares of length equal to 10cm. 


Finally, we recall that Algorithm~\ref{Second_alg} is considered for these simulations.

\subsection{Results}

\begin{figure}[t]
    \centering
    \includegraphics[scale=0.95]{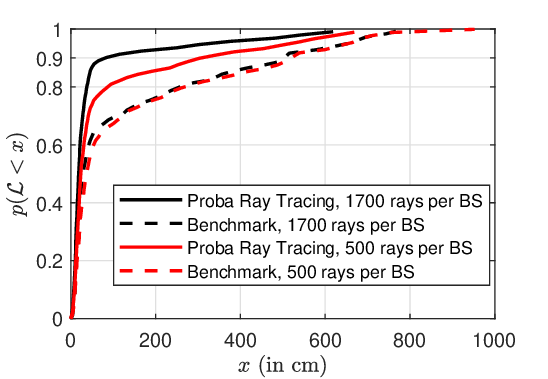}
    \caption{CDF of the positioning error with different number of rays both with the proposed method and the benchmark method. The standard deviation of the AoA measurement error is $\sigma =0.5$ degree.}
    \label{fig:simu_results_main}
\end{figure}

We show the cumulative density function (CDF) of the positioning error.
It is obtained as follows: For one position, we sample several AoA errors according to \eqref{equ_gaus_model}. For each sample, we compute the positioning error with both the proposed method and the benchmark method. We repeat the process for several positions.
Moreover,  we use the $Q\left(0.9\right)$ value (such that $90\%$ of the errors are under $Q\left(0.9\right)$) to compare the positioning performance of the algorithms. 
The positioning error is denoted by $\mathcal{L} = ||\hat{x}_k - x_k||$, where $x_k$ is the true position and $\hat{x}_k$ (center of $\mathcal{X}_k$) the estimated position.

We investigate the impact of the number of rays as well as the impact of the standard deviation of the measurement error~$\sigma$.

The results with different number of rays are provided on Figure~\ref{fig:simu_results_main}. The standard deviation of the angle error is $\sigma =0.5$ degree.
First, we see with the solid lines that the number of rays has a significant impact on the positioning error. The $Q(0.9)$ value decreases from 4m down to 50cm when the number of rays launched per BS increases from 500 to 1700. Note that the positioning performance with the low number of rays could probably be improved by increasing the size of the squares, as discussed in the previous section. We let this optimization for a future work. Second, we observe that the proposed method offers significant improvement compared to the benchmark method. The $Q(0.9)$ values decreases from 5m down to 50cm with 1700 rays. 

The accuracy achieved by the proposed methods with $\sigma=0.5$ is similar (even slightly better) to the fingerprint methods based on the channel impulse response (CIR) and neural networks. In \cite{TdocSumOct}\cite{Chatelier2023}, $Q(0.9)$ values around 1m are reported when neural networks are trained on dataset with inter-position spacing of 0.3m. Note that this latter performance is achieved assuming perfect estimation of the CIR while we consider measurement errors.

\begin{figure}[t]
    \centering
    \includegraphics[scale=0.95]{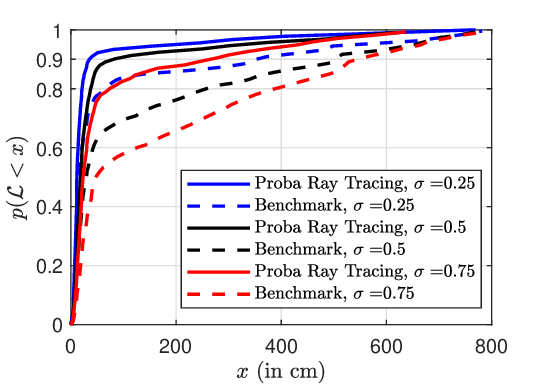}
    \caption{CDF of the positioning error with different values of standard deviation $\sigma$  of the measurement error both with the proposed method and the benchmark method. The number of rays per BS is 1700.}
    \label{fig:simu_results_sigma}
\end{figure}

The results with different standard deviation of the measurement error $\sigma$ are provided on Figure~\ref{fig:simu_results_sigma}. 
Unsurprisingly, the performance of both techniques decreases with a higher $\sigma$. 
Nevertheless, we observe that the proposed probabilistic method maintains a clear advantage over the benchmark method, especially for moderate $Q$ values. Thus, the method is relevant for several noise levels.

\section{Conclusions}

In this paper, we addressed the positioning problem based on AoA measurements in a NLoS environment.
A reverse ray-tracing solution, using a digital twin of the scene, is considered.
We proposed to take into account the statistics of the AoA measurement errors in a Bayesian manner to improve the reverse ray-tracing algorithm.
Simulation results show great improvements with the proposed method compared to the benchmark reverse ray-tracing method, where the statistics of the measurements are not taken into account in a Bayesian manner. 

\section{Appendix}
\label{App_1}
\begin{algorithm}
\caption{Uniform method to estimate the position via reverse ray tracing}
\label{first_alg}
\textbf{Inputs:} AoA measurements $\textbf{y}$, AoA statistics $p(\boldsymbol{\theta}|\textbf{y})$, and 3D model of environment (digital twin). \\
\begin{algorithmic}[1]
\STATE	For each of the $n$ BS, discretize uniformly $\mathcal{C}$ to have $l$ angles. 
\STATE	Launch the $n\ \times l$ rays in the discretized angle directions.
\FOR{ each distinct $\mathcal{X}_k$ in the scene}
\STATE	Collect all the rays crossing $\mathcal{X}_k$.
\STATE 	For all $m$ possible combinations of $n$ rays (where each of the $n$ rays comes from a different BS), compute the corresponding $\alpha_j=\prod_{i=1}^{n}{p\left(\theta_i\middle| y_i\right)}$. 
\STATE  	Compute $\beta_k=\sum_{j=1}^{m}\alpha_j$ to obtain \eqref{eq_proba_prod}.
\ENDFOR
\STATE 	The square with the highest probability is $\mathcal{X}_k$ where ${k=\text{arg \ max}_k\beta}_k$.
\end{algorithmic}
\end{algorithm}



\begin{thebibliography}{99}

\bibitem{Chatelier2023} B. Chatelier et al., ``Influence of Dataset Parameters on the Performance of Direct UE Positioning via Deep Learning,"  European Conference on Networks and Communications (EuCNC), June 2023. 

\bibitem{Deng2016} S. Deng, G. R. MacCartney, and T. S. Rappaport, ``Indoor and Outdoor 5G Diffraction Measurements and Models at 10, 20, and 26 GHz," NYU WIRELESS TR 2016-001, May. 2016.

\bibitem{Degli2014} V. Degli-Esposti et al., ``Ray-Tracing-Based mm-Wave Beamforming Assessment," IEEE Access, vol. 2, pp. 1314–1325, 2014.

\bibitem{ESPRIT} R. Roy and T. Kailath, ``ESPRIT - Estimation of Signal Parameters Via Rotational Invariance Techniques," IEEE Transactions on Acoustics, Speech, and Signal Processing, vol. 37, no. 7, 1989.

\bibitem{Kikuchi2006} S. Kikuchi, A. Sano, and H. Tsuji, ``Blind Mobile Positioning in Urban Environment Based on Ray-Tracing Analysis," EURASIP Journal on Applied Signal Processing, Jan. 2006.

\bibitem{Kong2006}  F. Kong, N. Zheng, G. Liu, and X. Ren, ``Position Orientating PathReverse Ray Tracing Algorithm Based on Cluster Analysis," 2016 IEEE SNPD, June 2016.

\bibitem{DS} L. C. Godara, ``Application of antenna arrays to mobile communications. II. Beam-forming and direction-of-arrival considerations," in Proceedings of the IEEE, vol. 85, no. 8, pp. 1195-1245, Aug. 1997.

\bibitem{MVDR} S. S.  Haykin, ``Adaptive filter theory." Pearson Education India, 2002.

\bibitem{Larew2013} S. G. Larew, T. A. Thomas, M. Cudak, and A. Ghosh, ``Air interface design and ray tracing study for 5G millimeter wave communications," IEEE Globecom Workshops, pp. 117–122, Dec 2013.

\bibitem{Lecci2020} M. Lecci et al., ``Simplified Ray Tracing for the Millimeter Wave Channel: A Performance Evaluation,"
2020 Information Theory and Applications Workshop (ITA), Feb. 2020.

\bibitem{Liu2014}  D. Liu, K. Liu, Y. Ma, and J. Yu, ``Joint TOA and DOA Localization in Indoor Environment Using Virtual Stations," IEEE communications letters, vol. 18, no. 8, Aug. 2014.

\bibitem{MUSIC} R. Schmidt, ``Multiple emitter location and signal parameter estimation,"  IEEE Transactions on Antennas and Propagation, vol. 34, no. 3, pp. 276-280, March 1986.


\bibitem{Wielandt2017} S. Wielandt and L. De Strycker, “Indoor Multipath Assisted Angle of Arrival Localization,” sensors, Nov. 2017.

\bibitem{5G_ACIA} 5G-ACIA RP-181521, ``LS on Channel Model for Indoor Industrial Scenarios," Meeting number 81,
Gold Coast, AU, Sept 2018. Available at: \url{https://www.3gpp.org/ftp/TSG_RAN/TSG_RAN/TSGR_81/Docs/}

\bibitem{TdocSumOct} 3GPP R1-2210652, ``Final Summary of Evaluation on AI/ML for positioning accuracy enhancement," Meeting number 110bis-e, e-Meeting, Oct. 2022.


\end{thebibliography}
\end{document}